\documentclass[journal=Langmuir,manuscript=article]{achemso}

\usepackage{chemformula} % Formula subscripts using \ch{}
\usepackage[T1]{fontenc} % Use modern font encodings

\author{Letizia Ferbel}
\email{letizia.ferbel@sns.it}
\phone{+39 3466079369}
\author{Stefano Veronesi}
\author{Stefan Heun}
\affiliation{NEST, Istituto Nanoscienze-CNR and Scuola Normale Superiore, Piazza San Silvestro 12, 56127 Pisa, Italy}

\title[]{Rb-induced ($3\times1$) and ($6\times1$) reconstructions on Si(111)-($7\times7$): A LEED and STM study}

\abbreviations{STM,LEED}

\begin{document}

\begin{abstract}
We present a study of the rubidium adsorption on the Si(111)-($7\times7$) surface and the related Rb-induced reconstructions as a function of deposition temperature and Rb-coverage via scanning tunneling microscopy (STM) and low-energy electron diffraction (LEED). Sample analysis via LEED allowed to observe for the first time a Rb/Si(111)-($6\times1$) reconstruction. The STM image analysis allowed to obtain the first real space characterization of the Rb/Si(111)-($3\times1$) surface. In addition, STM provided a direct and local information on the surface arrangement as well as further insights on the interaction between Si and Rb atoms and on the growth dynamics. 
\end{abstract}

\section{Introduction}
Reaction and interaction between metal atoms and semiconductor surfaces are widely studied \cite{christmann2006adsorption}. In particular, the rubidium-silicon system finds application in a large variety of fields. A knowledge of the structural, interfacial, chemical, and electronic properties of the system is thus foremost.

Rubidum, and in general alkali metals (AMs), have simple electronic structure, strong atomic transitions, and inherently high power density due to low ionization energy thus high reactivity. \cite{rumble2020crc}. Their reactivity, however, makes the processing, storage and use of AMs in a controlled manner difficult. The growing interest in developing quantum integrated systems sets the focus on miniaturized oxidation-free rubidium cells \cite{Knapkiewicz2018, Ruyack_2015, Sebbag2021}. These cells allow for better scalability and integration in nano-circuits, where commercial getters are not employable. They make use of non-magnetic materials (such as Si), and need to provide low amounts of pure Rb. Among the applications of these miniaturized Rb cells are highly sensitive optical magnetometries, atomic sensors \cite{Sebbag2021}, miniaturized atomic clocks \cite{Maurice:20}, quantum computing \cite{Saffman_2016}, and controllable combustion \cite{Ruyack_2015}.

Rubidium on Si has been demonstrated to improve the thermodynamical stability of multilayer Mo/Si mirrors, which are now the most viable technology to construct the optical systems for extreme-ultraviolet lithography \cite{SAEDI2020144951}.

The importance of AMs adsorbed on silicon finds also application in the field of catalysis. It has been reported that AMs can either promote oxidation of Si\cite{Soukiassian1986SiO2SiIF, PhysRevB.39.12775} or passivate the silicon surface depending on the alkali-induced silicon surface reconstruction \cite{PhysRevB.44.3222, NELSON1997365}. Catalytic oxidation using AMs finds its application in the formation of the SiO$_2$-Si interface. Standard silicon oxidation process requires very high temperatures which are not always compatible with device fabrication processes. The advantage of the AM as catalyst is that it can be removed by a moderate thermal annealing, thus with low power cosumption and the systems results contaminations-free. Certain types of AM-induced reconstructions of the Si surface, such as the ($3\times1$) reconstruction, have been reported to passivate the surface against oxidation. This also finds a large variety of applications, for example in solar cells \cite{GLUNZ2018260}. 

In the past, overlayers on semiconductor surfaces have attracted great attention for the study of metal-semiconductor interactions and interface formation as well as for metallization processes. Due to their simple electronic structure, AM overlayers on semiconductor surfaces represent simple model systems for the more general case of metal-semiconductor studies \cite{1989, doi:10.1142/S0218625X95000480}. Nowadays, the research focus has been set on ordered nanostructures, stimulated by their great potential for next generation devices. To control the nanostructure dimensions, spatial distribution and uniformity, knowledge of the mechanisms for their spontaneous formation is crucial. In particular, one-dimensional (1D) metal atomic chain systems on semiconductor surfaces have attracted substantial attention because of their intriguing electronic properties and potential applications in atomic scale devices \cite{Battaglia_2008, Do2018}.

Various self-assembled arrays of metal atomic chains on semiconductors have been observed at submonolayer metal coverages, and a ($3\times1$) phase has been frequently observed on the Si(111) surface. This reconstruction is formed by all AM (Li, Na, K, Rb, Cs) \cite{DAIMON1985320, PhysRevLett.69.1419, Fan1990, Wan1992, Paggel1993, JP271111, doi:10.1116/1.585546, BAKHTIZIN1995347, doi:10.1116/1.587700, OKUDA1994105, PhysRevB.50.1725, PhysRevB.52.5813, PhysRevB.52.1995}, monovalent Ag \cite{PhysRevB.65.045305, PhysRevB.66.161101}, alkaline earth metals (Mg, Ca, Sr, Ba) \cite{QUINN1991L307, SEKIGUCHI2001148, SARANIN200087,WEITERING1996L271}, and rare earth metals (Yb, Sm) \cite{PhysRevB.48.11014, PhysRevB.47.9663, PhysRevB.75.205420}. Some similarities as well as differences exist in the appearance of the reconstruction induced by various adsorbates and in particular among different AMs. Despite more than two decades of experimental efforts on these surfaces, some issues still need to be solved. 

\begin{figure}[!htb]
\centering
\includegraphics[width=0.6\linewidth]{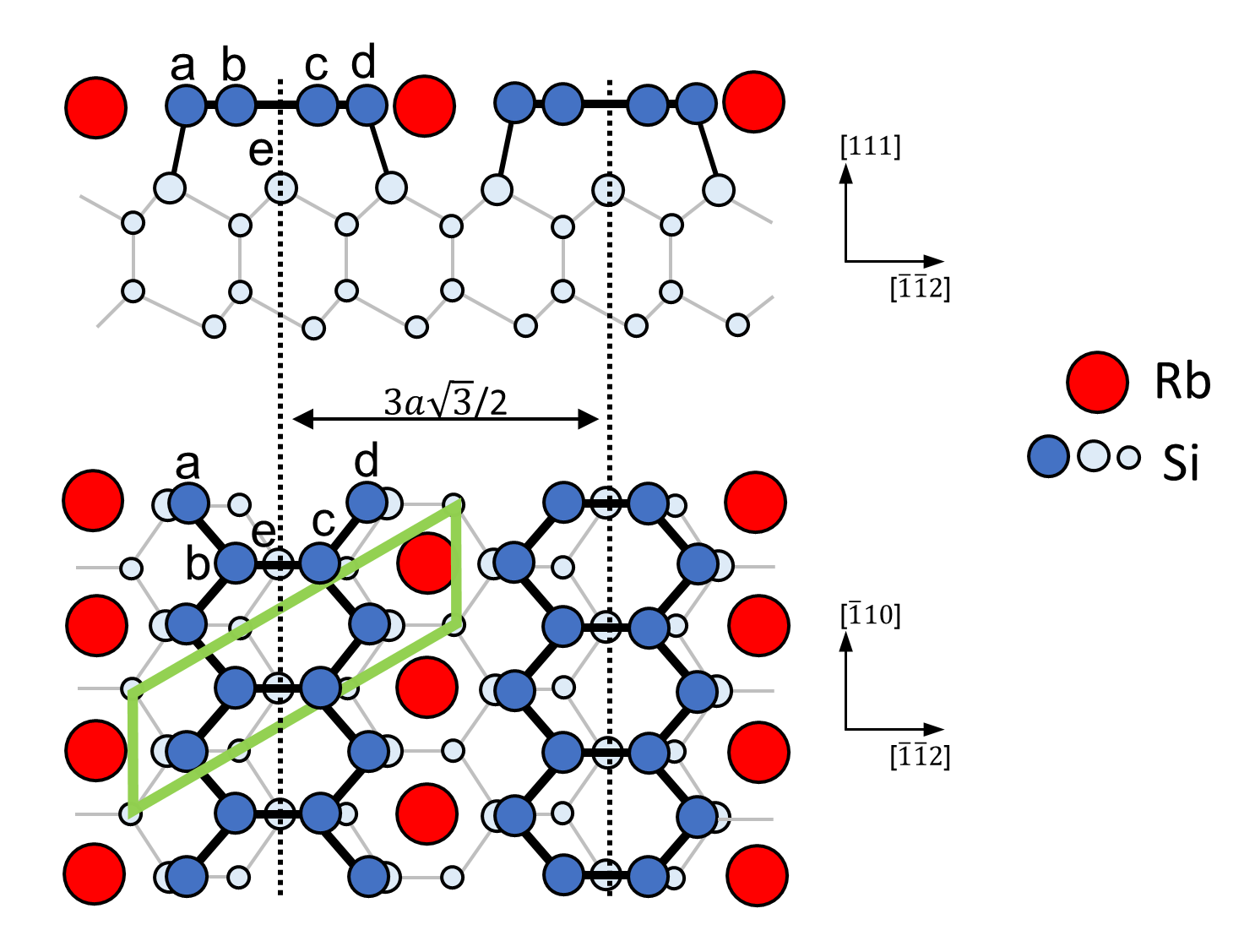}
\caption{\label{Fig1} Honeycomb  chain-channel (HCC) structure of the Si(111)-($3\times1$) surface with a Rb coverage of 1/3~ML. Red circles are Rb atoms. All other circles are Si atoms (dark blue, surface atoms, and light blue, bulk atoms). The ($3\times1$) surface unit cell is shown as green solid line.}
\end{figure}

The surface geometry of the ($3\times1$) reconstruction has been studied both experimentally and theoretically, and many structural models have been proposed. Among the proposed models, the honeycomb-chain-channel (HCC) \cite{PhysRevLett.80.1678,PhysRevLett.80.3980,PhysRevLett.81.2296} is nowadays widely accepted. In the HCC model, shown in Fig.~\ref{Fig1}, ``honeycomb chains'' are formed by four inequivalent Si surface atoms. The inner atoms (\textit{b}, \textit{c} in Fig.~\ref{Fig1}), threefold coordinated in a planar configuration via Si double bonds (sp$^2$-like), interact only weakly with the Si atom below, indicated as \textit{e} in Fig.~\ref{Fig1}. The resulting $\pi$-bond is primarily responsible for the stability of the HCC model. The dangling bond on the outer atoms (\textit{a}, \textit{d} in Fig.~\ref{Fig1}), roughly tetrahedrally coordinated, is fully occupied due to the electron donation of the AM and to the electron distribution associated with the Si adatoms and the underlying atom (\textit{e} in Fig.~\ref{Fig1}). The AM atoms are fully ionized and sit in a channel formed by neighboring honeycomb chains and form parallel ordered one-dimensional rows in the [$\bar{1}10$] direction. 

In agreement with the HCC model, recent experimental work arrived at a consensus coverage of $1/3$~ML for the AM in the ($3\times1$) phase, where one monolayer (ML) refers to the Si atomic density in the Si(111)-($1\times1$) surface ($\text{1~ML} =7.8\times 10^{14}\text{ cm}^{-2}$) \cite{PhysRevB.58.7059, PhysRevB.54.10585, sia.1029}. STM analysis of the ($3\times1$) surface reveals that this phase is imaged as double zig-zag rows of silicon atoms running along the [$\bar{1}10$] direction, separated by a missing row, which accommodates the AMs. The zig-zag rows are spaced by $3a\sqrt{3}/2\sim9.98$~\AA, where $a = 3.84$ \AA{} is the Si(111)-($1\times1$) lattice constant \cite{Wan1992}. A semiconducting character of the ($3\times1$) surface has been reported for Li \cite{PhysRevB.54.10585}, Na \cite{PhysRevB.55.6762, PhysRevLett.69.1419}, and K \cite{PhysRevB.50.1725}. In addition, passivation against oxidation has been reported for the Na \cite{PhysRevB.44.3222} and K /Si(111)-($3\times1$) surface \cite{NELSON1997365}. In the case of Cs adsorbed on the Si(111)-($7\times7$), photoemission spectroscopy and LEED showed that at high Cs coverage the ($3\times1$) structure is transformed into a ($6\times1$) \cite{PhysRevB.52.1995, JP271111}. An AM/Si(111)-($6\times1$) reconstruction has only been reported for Cs so far. 

In the particular case of the Rb/Si(111) surface, the experimental evidence of the ($3\times1$) reconstruction is limited to diffraction and core level spectroscopy studies. The ($3\times1$) reconstruction was observed depositing Rb on a clean Si(111)-($7\times7$) surface at elevated temperatures \cite{DAIMON1985320}. Moreover, x-ray standing wave studies of samples obtained by RT deposition of Rb reported an increased order of the surface after an annealing at $300-350$\textcelsius{} \cite{PhysRevB.48.12023}, attributed to the formation of the ($3\times1$) phase. No evidence of the ($3\times1$) reconstruction was found in these studies after RT deposition, but rather a preferential adsorption of Rb on multiple sites of the ($7\times7$) surface was reported \cite{PhysRevB.46.13631, Castrucci1993, Rodrigues_1993, PhysRevB.51.1581}. The Si 2p core level spectrum of the Rb/Si(111)-($3\times1$) surface was well fitted by one bulk and three surface components \cite{OKUDA199789}, consistent with the HCC model \cite{PhysRevLett.81.2296}. The Rb 4p spectrum indicated a single adsorbate site on the ($3\times1$) surface, consistent with an alkali coverage of 1/3~ML \cite{OKUDA199789}. 

In this paper we investigate the adsorption of Rb on the Si(111)-($7\times7$) surface and the related Rb-induced reconstructions. In particular, we analyze their dependence on deposition temperature and Rb-coverage with surface sensitive techniques such as LEED and STM. We report the first real space images of a Rb induced ($3\times1$) reconstruction as well as the first observation of a Rb($6\times1$) reconstruction. Finally, we discuss the mechanisms of reconstruction, analyzing the importance of substrate temperature and suggest an explanation for the appearance of the ($6\times1$) structure.

\section{Experimental}

The Si(111) samples were cut from commercial silicon wafers with 500 $\mu$m thickness. The whole experiment (sample preparation and characterization) has been performed in an ultra-high vacuum (UHV) environment, with a base pressure better than $10^{-10}$~mbar. Clean and well reconstructed Si(111)-($7\times7$) surfaces were prepared in situ by heating the as-cut samples ($2\text{ mm}\times 8\text{ mm}$) to 1000\textcelsius, after an overnight outgassing at 600\textcelsius. This preparation method allows to obtain a Si(111)-($7\times7$) surface characterized by clean, parallel, and fairly regular ($\sim50$ nm large) flat terraces separated by steps $0.3-1.5$~nm high. Rb was evaporated from a commercial getter source (SAES Getters Inc.) with a rate of about 0.03 ML/min onto the Si(111)-($7\times7$) surface held either at room temperature or at elevated temperatures ranging from 300 to 400\textcelsius. Depositions were performed at constant evaporation rate,  monitored with a residual gas analyzer (RGA from SRS), on samples kept at constant temperature. Temperatures were measured with an infrared pyrometer with an uncertainty of $\pm 10$\textcelsius. LEED and STM characterization was performed at RT. The scanning tunneling microscope used to perform the experiments is a variable-temperature UHV STM from RHK Technology working at a base pressure of $3\times 10^{-11}$~mbar. The tips used to scan the samples are electrochemically etched tungsten tips that were first degassed in situ and then flashed by applying high voltage. All STM images were obtained in constant-current mode.

\section{Results and Discussion}
%%%
\begin{figure}[!b]
\includegraphics[width=\linewidth]{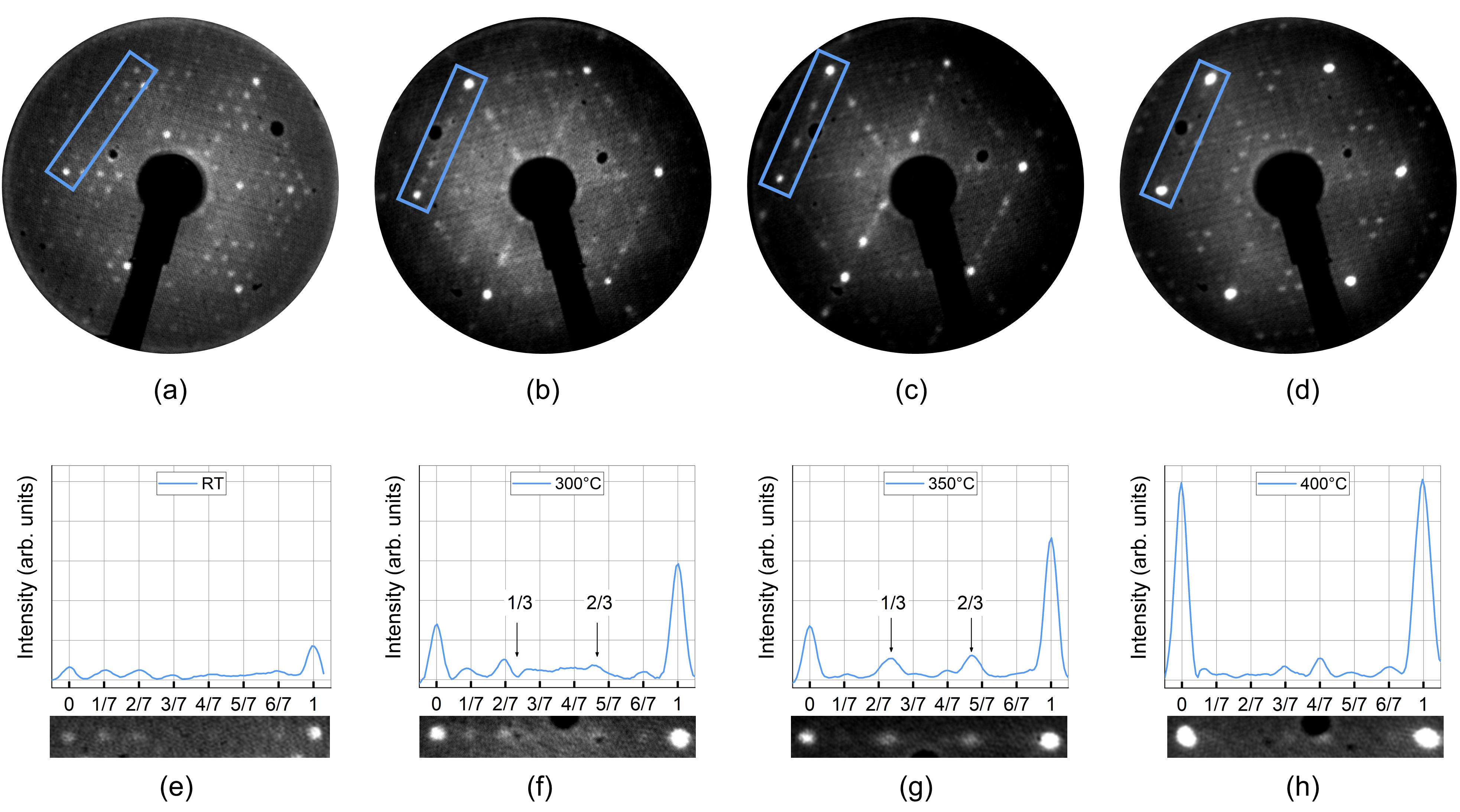}
\caption{\label{Fig2} LEED patterns of Rb/Si(111) surfaces after 9~min Rb deposition on the ($7\times7$) substrate held at (a) RT, (b) 300\textcelsius, (c) 350\textcelsius{}, and (d) 400\textcelsius. Beam energy: 47 eV. (e)-(h) Cross sections and cropped areas taken in the regions highlighted by the blue rectangles in (a)-(d). The intensity scale is the same for all cross sections.}
\end{figure}

We first report the results on the Rb adsorption behavior as a function of the deposition temperature. Figure~\ref{Fig2} shows the LEED patterns of Rb/Si(111) surfaces obtained after 9~minutes of Rb deposition at different temperatures on the clean ($7\times7$) surface. When Rb was deposited at RT, the LEED pattern (Fig.~\ref{Fig2}(a) and \ref{Fig2}(e)) retained a ($7\times7$) periodicity. However, the surface showed a diffraction pattern with a different intensity distribution compared to the original ($7\times7$) pattern observed on the clean silicon sample, underlined by the asymmetry of the ($1\times1$) spots otherwise symmetric at this beam energy, as well as highly suppressed spot intensities. When Rb was deposited at 300\textcelsius, the LEED pattern (Fig.~\ref{Fig2}(b) and \ref{Fig2}(f)) readily showed the presence of weak and broad spots arranged in a ($3\times1$) pattern, superimposed on a sharp pattern with ($7\times7$) periodicity. As the deposition temperature was increased, the ($3\times1$) pattern became more clear, reaching its maximum intensity with deposition temperatures around 350\textcelsius{} (see Fig.~\ref{Fig2}(c) and \ref{Fig2}(g)). Around this temperature, the ($3\times1$) spots became sharp and clearly distinguishable, while the ($7\times7$) spots decreased in intensity. Upon deposition at 400\textcelsius{} or higher, the LEED pattern showed again a ($7\times7$) periodicity (Fig.~\ref{Fig2}(d) and \ref{Fig2}(h)) and no evidence of a ($3\times1$) reconstruction. The intensity distribution of this diffraction pattern resembles that of the clean reconstructed silicon sample, in strong contrast to that obtained upon RT deposition. 
 
\begin{figure}[!hbt]
\includegraphics[width=0.5\linewidth]{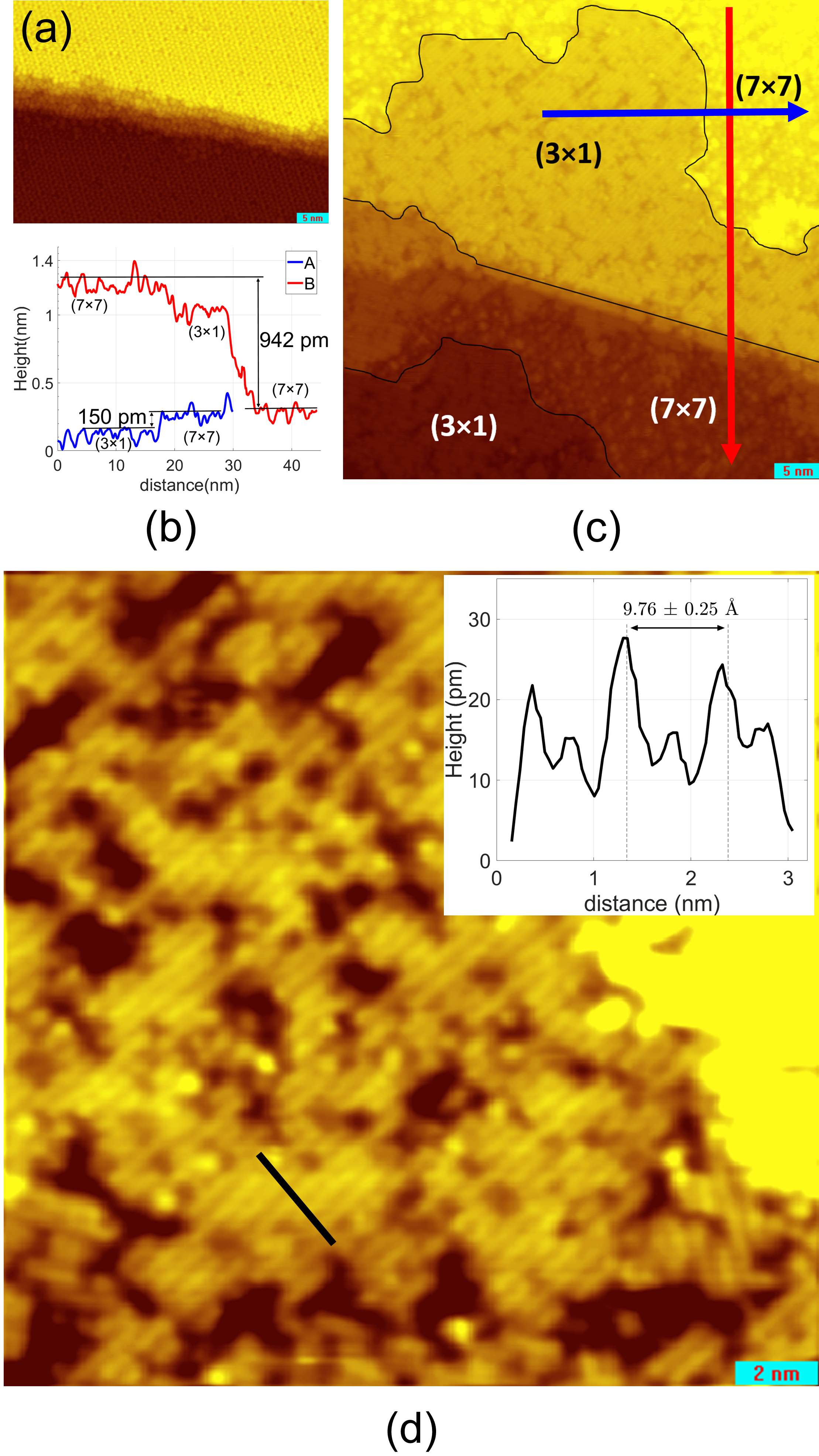}
\caption{\label{Fig3} STM images (a) of the clean ($7\times7$) surface and (c), (d) of the Rb/Si(111) surface obtained upon 9~min Rb deposition on the ($7\times7$) surface held at 350\textcelsius. Image parameters: $-2.5$ V, 20 pA, image size: (a) $50\text{ nm}\times20\text{ nm}$, (c) $50\text{ nm}\times50\text{ nm}$, (d) $20\text{ nm}\times20\text{ nm}$. (b): cross sections taken along the blue and red lines drawn in (c), reporting the measured height differences between ($3\times1$) and ($7\times7$) areas on the same terrace and between two different ($7\times7$) terraces, respectively. Inset in (d): cross section taken along the black line drawn in (c), reporting the measured periodicity of the row structure.}
\end{figure}

STM analysis of the samples shown in Fig.~\ref{Fig2} allowed to account for the amount of ($3\times1$) areas. Figure~\ref{Fig3}(c) shows a large scale STM image of a 9~min Rb-dosed sample (LEED pattern in Fig.~\ref{Fig2}(c)), in which a change in the terrace geometry can readily be seen compared to the pristine sample (in Fig.~\ref{Fig3}(a)). The flat ($7\times7$) terraces observed in the pristine sample are modified by the Rb adsorption into a two-level system, as highlighted by the blue cross section in Fig.~\ref{Fig3}(b). The upper level retains a ($7\times7$) periodicity, differing from the pristine surface only by the presence of extra atoms adsorbed on the surface. The lower level is occupied by a row structure, characteristic of the ($3\times1$) reconstruction. The red cross section in Fig.~\ref{Fig3}(b) shows that the step height between ($7\times7$) levels did not change upon Rb deposition, and its height is an integer multiple of $314$~pm, as expected. In addition, the areas characterized by the rows extend inwards from the step edges, as shown in Fig.~\ref{Fig3}(c). Moreover, the difference in silicon atoms density between the ($3\times1$) and the ($7\times7$) reconstructions suggests a high silicon mass transport in the transformation from ($7\times7$) to ($3\times1$), and thus the presence of excess silicon.

Figure~\ref{Fig3}(d) shows a zoom into the lower level with the row structure: the filled state STM image shows parallel double rows separated by a “missing row” with a measured distance between the double rows of 9.76$\pm$0.25 \AA. Some disorder in the ($3\times1$) areas is represented by defects like extra atoms adsorbed on the rows or vacancies within the rows. The geometry of these areas together with the very good agreement with the theoretical distance between the rows \cite{Wan1992} makes this the first real space evidence of a Rb-induced ($3 \times 1$) Si(111) reconstruction. 

From multiple STM scans of Rb/Si samples after 9~minutes of Rb deposition, the amount of ($3\times1$) areas resulted in 0\% in the case of RT and 400\textcelsius{} deposition, and 22\% and 65\% in the cases of deposition at 300\textcelsius{} and 350\textcelsius, respectively. In addition, in the case of RT deposition, we found a vast amount of Rb adsorbed on the ($7\times7$) surface in a disordered fashion, but preferentially on the faulted half of the unit cell (refer to Supplementary Material). In this case, after annealing the Rb/Si sample at 300\textcelsius, small patches of ($3\times1$) reconstruction on the surface were obtained (refer to Supplementary Material). On the contrary, in the case of Rb deposition at 400\textcelsius, only a small number of Rb atoms was found on the silicon surface, not allowing the ($3\times1$) to form even after annealing cycles. The aforementioned STM observations prove that at temperatures below 400\textcelsius, Rb atoms adsorb on the Si(111) surface, while around and above 400\textcelsius{} they do not, consistent with the presented LEED evidence. Upon deposition at 300-350\textcelsius, Rb atoms arrange to form a surface structure with ($3\times1$) periodicity, thus the temperature range for the formation of this reconstruction is at least 50\textcelsius. Since we found lowest intensity of the ($7\times7$) and highest intensity of the ($3\times1$) diffraction spots as well as the highest amount of ($3\times1$) surface areas upon deposition at 350\textcelsius, we conclude that 350\textcelsius{} is the optimal temperature for this reconstruction to form, and the following analysis will be centered on samples prepared at this deposition temperature.
 
\begin{figure}[!hbt]
\centering
 \includegraphics[width=0.5\linewidth]{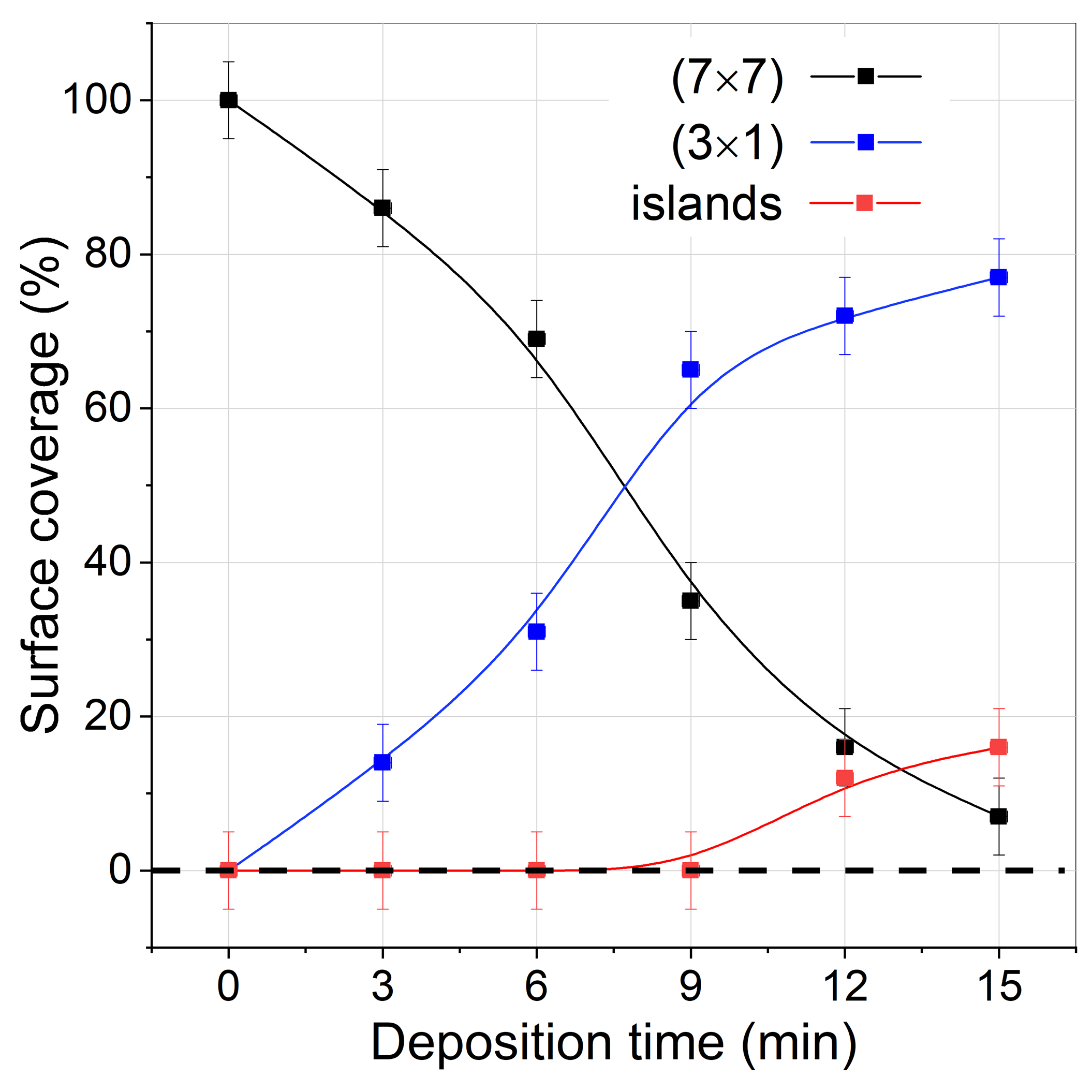} %
 \caption{\label{Fig4}  STM analysis of the samples surface: fraction of ($7\times7$) and ($3\times1$) areas as a function of the deposition time. The spline connecting the data points is shown as a guide to the eye. Deposition temperature 350\textcelsius.}
 \end{figure}

\begin{figure}[!htpb]
\centering
\includegraphics[width=0.5\linewidth]{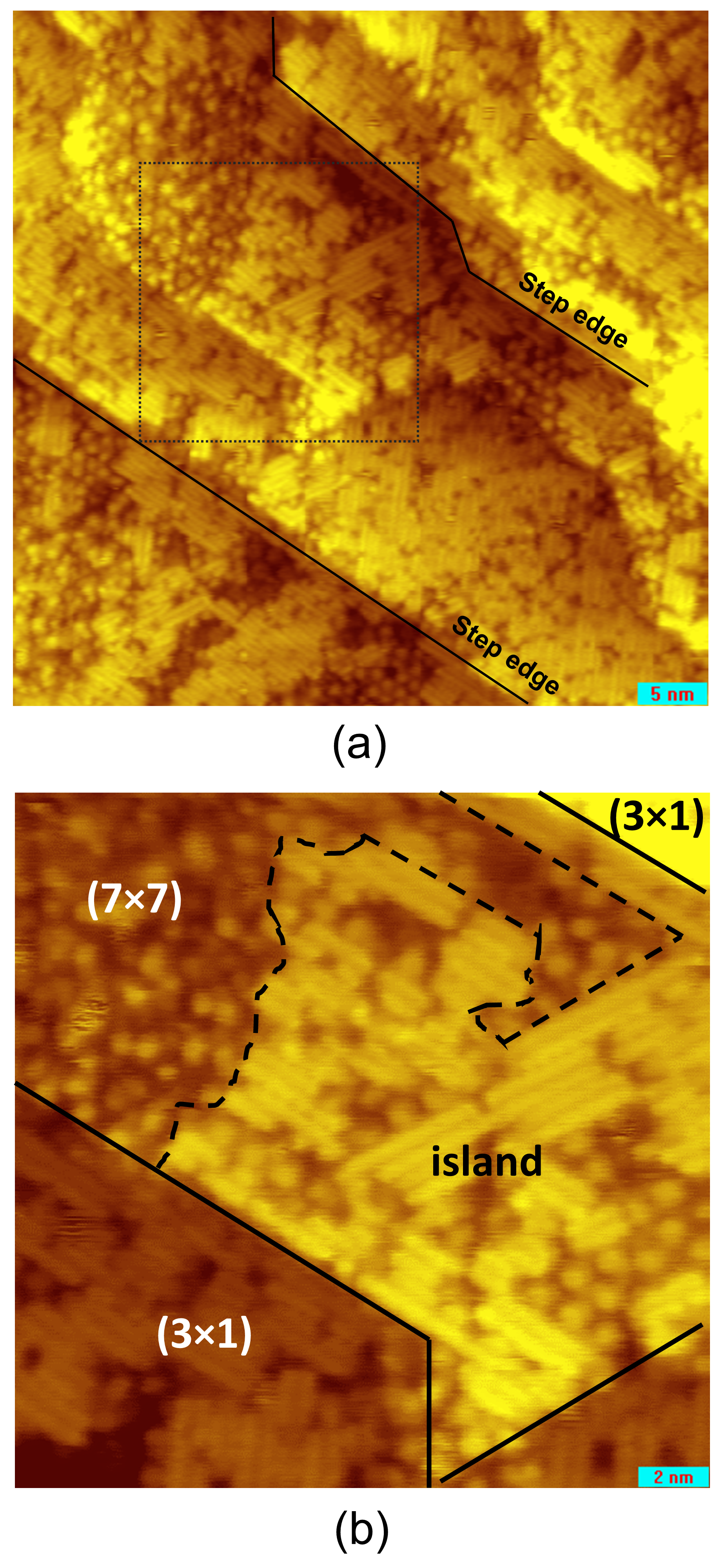}
\caption{\label{Fig5} STM images of the Rb/Si(111) surface obtained upon 12~min Rb deposition on the ($7\times7$) surface at 350\textcelsius. Image parameters: $-3.0$ V, $0.8$ nA. Image size: (a) $50\text{ nm}\times50\text{ nm}$, (b) $20\text{ nm}\times20\text{ nm}$. Image (b) was taken in the area marked in (a) by a dashed square.}
\end{figure}

The influence of Rb surface coverage on the reconstruction was studied, as well. Figure~\ref{Fig4} reports the results of STM analysis and shows the amount of different reconstructions found on the samples as a function of the Rb deposition time at 350\textcelsius. The plot in Fig.~\ref{Fig4} clearly shows that, in the studied deposition time range, with increasing deposition time the area retaining a ($7\times7$) reconstruction decreases. The presence of a ($3\times1$) phase was already revealed after 3~minutes of Rb deposition, both in the LEED patterns and STM images. With deposition time as low as 3~minutes, the terrace system started to transform into a two-level system, with coexistence of ($3\times1$) and ($7\times7$) phases on the same terrace, as reported in Fig.~\ref{Fig3}. Up to 9~minutes deposition time, an increase in deposition time led to an increase in the number of ($3\times1$) domains as well as an increase in their size, with preferential growth in the [$\bar{1}10$] direction. The two-level system was maintained, as well. After 12~minutes of deposition, we observe a slowdown in the growth of the ($3\times1$) areas composing the lower level of the terraces. At the same time, we observe the presence of islands on the upper ($7\times7$) level, characterized by a row structure on their surface, as reported in Fig.~\ref{Fig5}. In addition to these rows, between already formed rows we observe a reordering of atoms towards formation of a row structure, as visible in Fig.~\ref{Fig5}(b).

 \begin{figure}[!htpb]
\centering
\includegraphics[width=0.3\linewidth]{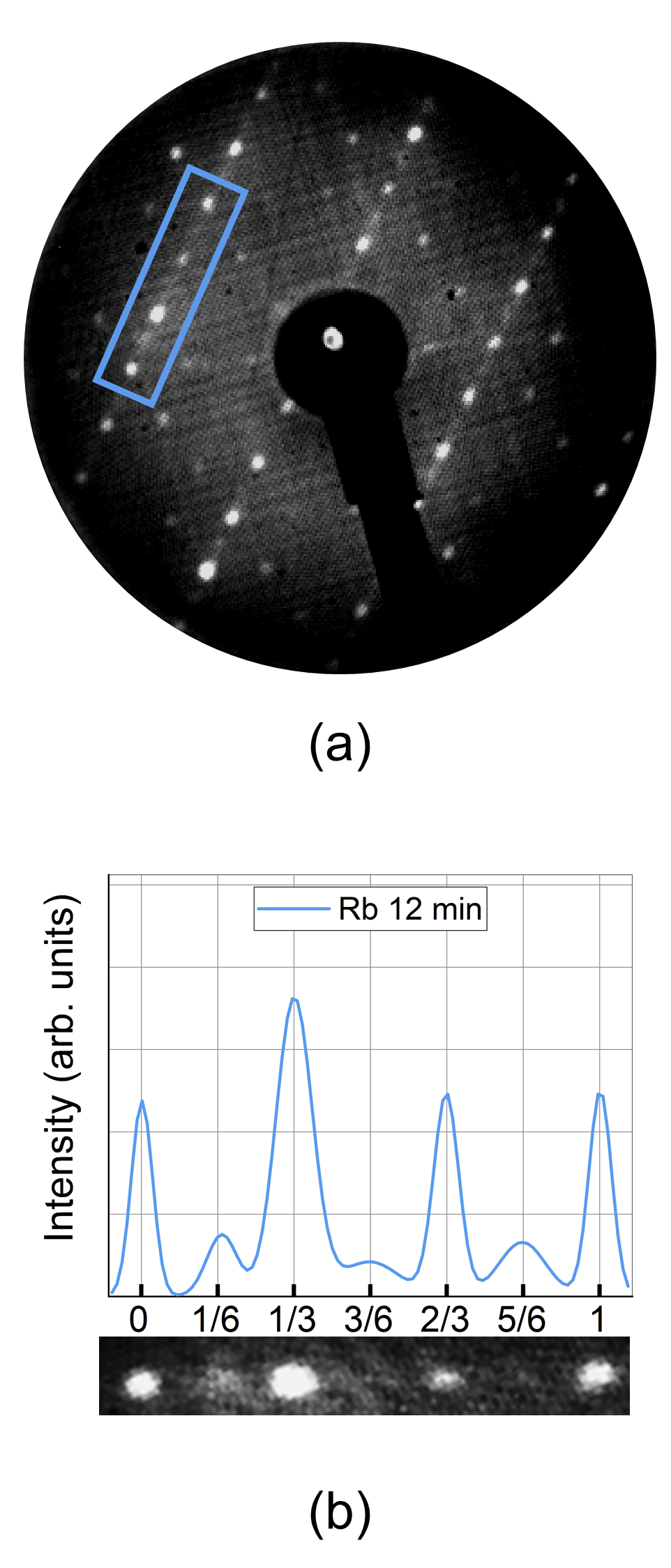}%
\caption{\label{Fig6} (a) LEED pattern of a Rb/Si(111) surface displaying ($3\times1$) and ($6\times1$) periodicity. Beam Energy: 76 eV. We choose this beam energy because the ($6\times1$) spots were more evident at this beam energy, even if the ($6\times1$) spots were already evident at 47 eV. (b) Cross section and cropped area taken in the region highlighted by the blue rectangle in the LEED pattern in (a). Sample obtained by dosing Rb for 12~min on the Si(111)-($7\times7$) surface held at 350\textcelsius{} during deposition.}
\end{figure} 

Interestingly, besides the ($3\times1$) spots, the LEED patterns obtained upon 12 and 15~min Rb deposition at 350\textcelsius{} revealed the presence of weak and broad spots arranged in a ($6\times1$) pattern (see Fig.~\ref{Fig6}). The weakness of the ($6\times1$) spots suggests that only a small fraction of the surface is arranged with this periodicity. Moreover, the broadening of the ($6\times1$) spots indicates the presence of small domains. It is to underline that this is the first time that a Rb($6\times1$) structure is reported. 

\begin{figure}[!tbhp]
\centering
\includegraphics[width=0.5\linewidth]{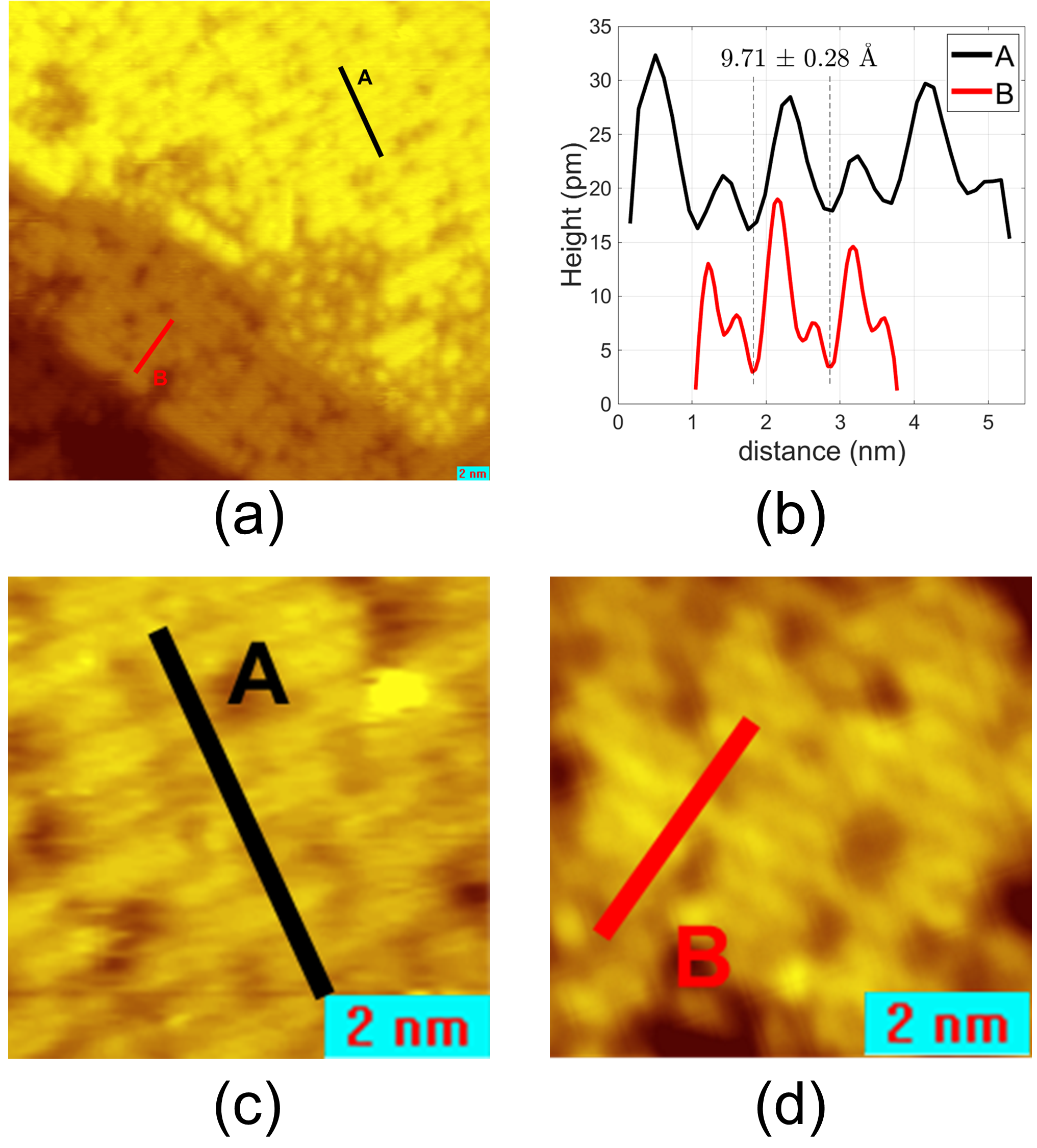}
\caption{\label{Fig7} STM image of the Rb/Si(111) surface obtained upon 12~min Rb deposition on the ($7\times7$) surface at 350\textcelsius. Image parameters: $-3.0$ V, $0.8$ nA. Image size: $30\text{ nm}\times30\text{ nm}$. (b) Cross sections along the black and red lines indicated in (a). To better distinguish the difference between the two curves a height offset of 15 pm has been added to the cross section labeled as A. (c) and (d) Zoom-in in the cross section region reported in (a) and (b). The ($3\times1$) periodicity is measured to be $9.71 \pm 0.28$~\AA{} and indicated in (b) by the two vertical dashed lines.}
 \end{figure}

Our STM results show that row structures could be identified both on the lower terrace level and on the islands grown on the upper ($7\times7$) terrace level, as highlighted in Fig.~\ref{Fig5}(b). However, for some of the islands on the ($7\times7$) terraces we observe a different height distribution between neighbouring rows (A, black cross section in Fig.~\ref{Fig7}) when compared to the height distribution of the ($3\times1$) areas (B, red cross section in Fig.~\ref{Fig7}). Despite the fact that single rows could not be resolved in the cross section taken on the island, we note that each peak in (A) corresponds to two neighbouring rows in (B). This suggests that two neighboring rows on the island are seen  merged together as a single row. Besides, there is an alternation in the height of these structures in (A), which is markedly different from (B) where each ($3\times1$) element is imaged with the same height. Thus, this alternation of height results in a doubled periodicity with respect to the ($3\times1$) structure, which means that these atoms are arranged in a ($6\times1$) structure.

Based on our experimental observations, we suggest for the reconstruction process the following model. A balance between adsorption, desorption, and diffusion processes must be reached in order for the reconstruction processes to happen. Indeed, we observed that the ($3\times1$) reconstruction appears after RT deposition only upon annealing, or when the substrate is held at elevated temperature during deposition, in a precise and narrow range of temperatures. Upon deposition at RT, Rb atoms have small diffusion length. Nevertheless, despite the low mobility, the system will try to reach the configuration of lowest energy. What we see is adsorption preferentially on the faulted half of the unit cell. When Rb is deposited at higher temperatures, the thermal energy will promote Rb diffusion on the surface and surface reactivity. If desorption can be neglected, the ($3\times1$) conversion process can start. On the other hand, STM shows that at 400\textcelsius{} and above the desorption process dominates, since little or no Rb atoms were seen adsorbed on the surface, and the reconstruction does not form.  

\begin{figure}[!tbh]
\centering
\includegraphics[width=0.5\linewidth]{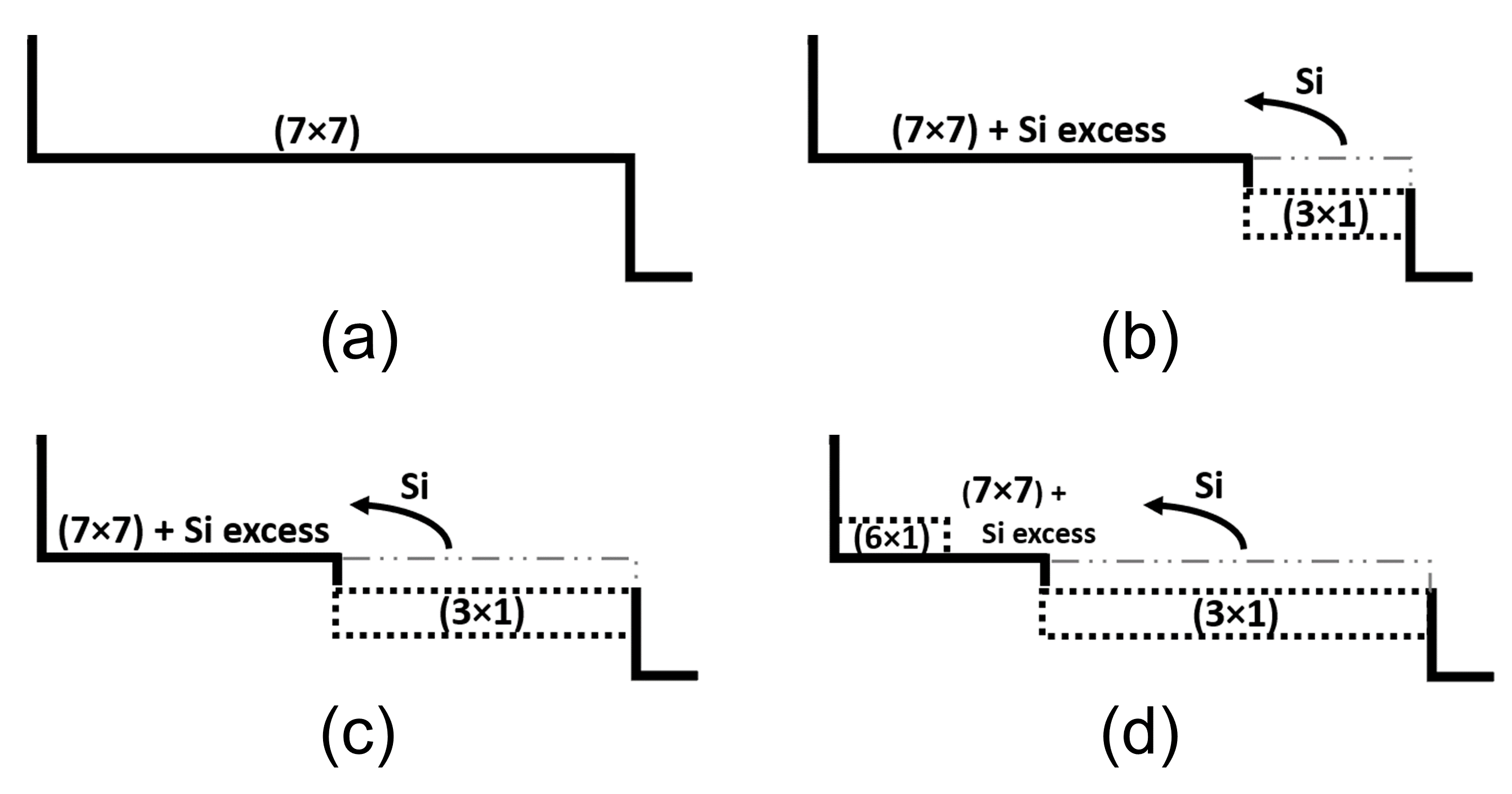}
\caption{\label{Fig8} Scheme of the surface reconstruction process upon Rb deposition. (a) pristine Si sample, (b)-(d) increasing Rb coverage.}%
\end{figure}

Supposing to be in the optimal temperature conditions for the reconstruction process to happen, then the deposited Rb atoms can migrate on the ($7\times7$) surface and react at suitable sites to form the ($3\times1$) structure. At the initial stage of formation, we find that suitable sites are at the step edges. These areas are characterized by silicon atoms having higher free energy given by the broken periodicity in the plane, thus an additional dangling bond. Rb reacting with Si atoms will progress towards the energy minimum, which is represented by the ($3\times1$) phase. As long as we deposit low-medium amounts of Rb, the Rb-Si reaction will take place predominantly at the step edges and proceed by enlarging row-structure domains inward from the step edge (as sketched in Fig.~\ref{Fig8}(b) and (c)). Since the ($3\times1$) phase is characterized by a smaller Si atom density compared to the ($7\times7$) structure ($\theta^{(7\times7)}_{Si}=2.08$~ML and $\theta^{(3\times1)}_{Si}=1.33$~ML \cite{PhysRevB.58.3545}), there will be an excess of Si atoms. The Si surplus is ejected onto the surrounding ($7\times7$) region (as shown in Fig.~\ref{Fig8}(b)). This suggests that the excess atoms observed on the ($7\times7$) surface are indeed Si atoms. 

However, when the excess of silicon becomes important, new favored reaction sites will be formed. Deposited Rb will also react with the excess silicon and form stable reconstructed islands on the ($7\times7$) areas, characterized by rows (as sketched in Fig.~\ref{Fig8}(d)). Besides, the formation of islands as well as the increase in the size of ($3\times1$) areas will lead to an important accumulation of compressive strain on the surface.

Thus, the appearance of a ($6\times1$) reconstruction for high Rb coverages could be explained in terms of strain release and large Rb atomic size. Strain in the islands would more easily lead to changes in the distance. Conversely, the ($3\times1$) reconstruction on the lower terrace levels is more strongly bound to the Si bulk, reconstructed on larger areas, and thus more stable. This might explain why we observe the ($6\times1$) only on the islands.

Moreover, the ($6\times1$) phase with doubled periodicity could be attributed to the large atomic size of Rb atoms ($r_a = 220 \pm 9$~pm) and the fact that the Rb nearest neighbor distance (4.84 \AA) is larger than the Si(111)-($1\times1$) lattice spacing (3.84~\AA) \cite{app5}. An increased interaction between the two $\pi$-bonded chains due to the large Rb atoms will cause the two zigzag chain rows to accumulate compressive strain, which can be released by a slight shift between alternating rows to produce the ($6\times1$) structure, consistent with the results shown in Fig.~\ref{Fig7} and results reported for Cs adsorbed on Si(111) \cite{JP271111, PhysRevB.52.1995}.

\section{Conclusions}
The dependence of temperature and deposition time on Rb induced Si(111)-($7\times7$) reconstructions was studied. We noted that if the deposition temperature is too low (e.g. RT), the ($3\times1$) reconstruction will not form. Rb will instead adsorb on the surface preferentially on the faulted half of the ($7\times7$) unit cell to minimize the surface energy. On the contrary, if the deposition temperature is too high (above 400\textcelsius), Rb atoms will not even stick to the surface, but desorb. Our experimental data suggest 350\textcelsius{} as the optimal temperature for the formation of the ($3\times1$) reconstruction. We show that a range of at least 50\textcelsius{} wide of suitable temperatures exists for the reconstruction to form. Furthermore, we analyzed the effect of Rb coverage. We obtained the first LEED evidence of a Rb($6\times1$) as well as the first real space characterization of the Rb induced ($3\times1$) on the Si(111)-($7\times7$) surface. We suggest a model to explain the growth of these new phases. We analyze the influence of the substrate temperature in terms of diffusion, adsorption and desorption processes and explain the presence of preferential adsorption sites on the surface. Finally, we expose the reasons for the formation of the ($6\times1$) as a result of a high amount of deposited Rb in terms of the large Rb atomic size and accumulated strain release from the surface. 

Our work allows for a better understanding of Rb interactions with Si. In particular, we discuss in detail how Rb self-assembles into 1D-atomic chains on the Si(111) surface. We provide a detailed recipe to obtain it. The electronic industry is still based on silicon technologies, thus the reduction in dimensionality and the discovery of systems with new properties make studies such as 1D-atomic metal chains on silicon interesting and foremost. Moreover, this Rb-silicon system represents an architecture that might be used to produce in a quite simple way miniaturized Rb cells of extremely pure Rb adsorbed on a non magnetic substrate, and atomically calibrated thanks to the formation of ordered Rb structure on Si(111). In addition, which usage requires small power consumption. Reduction in size, thus integration in nano-circuits, has highlighted the attention on such systems.

\begin{suppinfo}
STM and LEED study of RT deposition of Rb on the Si(111)-($7\times7$) surface, and after an annealing cycle at 300\textcelsius. Rb adsorption behavior and surface induced reconstruction are presented.
\end{suppinfo}

\bibliography{bibliography}

\end{document}